\documentclass[12pt]{iopart}
\usepackage{iopams,mathrsfs,graphicx,hyperref,color}
\expandafter\let\csname equation*\endcsname\relax
\expandafter\let\csname endequation*\endcsname\relax
\newcommand{\gae}{\lower 2pt \hbox{$\, \buildrel {\scriptstyle >}\over {\scriptstyle
\sim}\,$}}
\newcommand{\lae}{\lower 2pt \hbox{$\, \buildrel {\scriptstyle <}\over {\scriptstyle
\sim}\,$}}
\usepackage{amsmath}

\begin{document}

\title[Exact correlations in the nonequilibrium stationary state of the noisy
Kuramoto model]{Exact correlations in the nonequilibrium stationary state of the noisy Kuramoto model}

\author{Debraj Das and Shamik Gupta}

\address{Department of Physics, Ramakrishna Mission Vivekananda
University, Belur Math, Howrah 711202, India}
\ead{debraj.das@rkmvu.ac.in, shamik.gupta@rkmvu.ac.in}
\vspace{10pt}
\begin{indented}
\item[]June 2018
\end{indented}

\begin{abstract}
We obtain exact results on autocorrelation of the order parameter in
the nonequilibrium stationary state of a paradigmatic model of
spontaneous collective synchronization, the Kuramoto model
of coupled oscillators, evolving in presence of Gaussian, white noise.
The method relies on an exact mapping of the stationary-state dynamics
of the model in the thermodynamic limit to the noisy dynamics of a single, non-uniform oscillator, and allows to obtain
besides the Kuramoto model the autocorrelation in the
equilibrium stationary state of a related model of long-range
interactions, the Brownian mean-field model. Both the models show a
phase transition between a synchronized and an incoherent phase at a
critical value of the noise strength. Our results indicate that in the
two phases as well as at the critical point, the autocorrelation for
both the model decays as an exponential with a rate that increases
continuously with the noise strength. 
\end{abstract}

\vspace{2pc}
\section{Introduction}
\label{sec:intro}

Characterizing the stationary state of a many-body interacting system
evolving according to a given dynamics constitutes one
of the primary objectives of statistical mechanics \cite{Huang:1987}. Complexity in the
computation often stems from the many-body nature of the dynamics, and
is further enhanced if the stationary state is out of equilibrium
\cite{Livi:2017}.
Indeed, the phase-space distribution in an equilibrium stationary state
is given unequivocally by the Gibbs-Boltzmann weight independent of the
underlying dynamics leading to its attainment, while that in a
nonequilibrium stationary state does not have a universal form but has
to be obtained from an explicit consideration of the dynamics. While the
phase-space distribution is a characterization of a one-time snapshot of
the possible values of the dynamical variables in the stationary state, it is obviously of interest to consider how
do the values at one time relate to those at another time. A measure of
similarity of the values of dynamical variables at two different times
as a function of the time lag between them is given by the
autocorrelation function, which thereby provides valuable insights into
the underlying dynamics. Autocorrelations in equilibrium may be
deduced from the response of the system to small external perturbations 
by invoking the framework of the linear response theory \cite{Kubo:1966}. By contrast, there is as yet
no general procedure that allows for evaluation of autocorrelation in generic nonequilibrium stationary states, thus
warranting the need to study model systems for which explicit formulas
may be derived for autocorrelation.

In this work, we address the issue of obtaining exact results on the autocorrelation of a paradigmatic model showing
spontaneous order in a nonequilibrium stationary state, the Kuramoto
model. The model serves as a minimal framework to study the
phenomenon of spontaneous synchronization among a population of coupled
oscillating units of diverse natural frequencies \cite{Pikovsky:2001}. Spontaneous synchrony is commonly
observed in nature, e.g., in yeast cell suspensions \cite{Bier:2000}, flashing
fireflies \cite{Buck:1988}, arrays of Josephson junctions
\cite{Wiesenfeld:1998}, laser arrays \cite{Hirosawa:2013},
power-grid networks \cite{Rohden:2012}, and others. The Kuramoto model involves a
set of limit-cycle oscillators of distributed frequencies that are coupled
all-to-all through an interaction that depends sinusoidally on the
difference of the phases between the oscillators
\cite{Kuramoto:1984,Strogatz:2000,Acebron:2005,Gupta:2014-2,Gupta:2018}.
The noisy Kuramoto model considers in addition the fact that the
frequencies of the oscillators need not be constant in time but may have
stochastic fluctuations in time. Denoting
by $\theta_i \in [0,2\pi);~i=1,2,\ldots,N$ the phase of the $i$-th oscillator in a group
of $N$ oscillators, the dynamics of the model is given by a set of $N$
coupled Langevin equations of the form \cite{Sakaguchi:1988}
\begin{equation}
\frac{{\rm d}\theta_i}{{\rm d}t}=\omega_i + \frac{K}{N}\sum_{j=1}^N \sin
(\theta_j-\theta_i)+\eta_i(t).
\label{eq:Kuramoto-eom}
\end{equation}
Here, $K\ge 0$ is the coupling constant, $\omega_i \in [-\infty,\infty]$ is the natural
frequency of the $i$-th oscillator, while the noise $\eta_i(t)$
satisfies 
\begin{equation}
\langle \eta_i(t)\rangle=0,~\langle
\eta_i(t)\eta_j(t')\rangle=2D\delta_{ij}\delta(t-t'),
\end{equation}
with $D>0$ characterizing the strength of the noise and angular brackets
denoting averaging over noise realizations. The frequencies
$\{\omega_i\}_{1 \le i \le N}$ denote a set of quenched disordered random
variables distributed according to a common distribution $g(\omega)$,
with the latter obeying the normalization $\int_{-\infty}^\infty {\rm
d}\omega~g(\omega)=1$. As is often the case with most studies of the
Kuramoto model, we will consider $g(\omega)$ to be a unimodal distribution
with a non-compact support, that is, one which is symmetric about its mean $\langle \omega \rangle \equiv \int_{-\infty}^\infty {\rm
d}\omega~\omega g(\omega)$, and which decreases monotonically and
continuously to zero with increasing $]\omega-\langle \omega \rangle|$.

Now, it is evident from Eq. (\ref{eq:Kuramoto-eom}) that
the dynamics is invariant under the Galilean transformation $\theta_i \to
\theta_i + \langle \omega \rangle t ~\forall~i$. In the particular case when the
frequency term on the right hand side of the dynamics
(\ref{eq:Kuramoto-eom}) is absent (i.e., $\omega_i=0~\forall~i$), or, when all
the oscillators have the same frequency (equal to $\omega_0$, say, so
that $g(\omega)=\delta(\omega-\omega_0)$) and one observes the dynamics
(\ref{eq:Kuramoto-eom}) in a frame rotating uniformly with frequency
$\omega_0$ with respect to an inertial frame, the corresponding equations of motion
are given by
\begin{equation}
\frac{{\rm d}\theta_i}{{\rm d}t}=\frac{K}{N}\sum_{j=1}^N \sin
(\theta_j-\theta_i)+\eta_i(t).
\label{eq:BMF-eom}
\end{equation}
The above equations of motion constitute the so-called Brownian
mean-field (BMF) model \cite{Chavanis:2014}, and mimic the canonical
ensemble dynamics, namely, the overdamped dynamics in contact with a
heat bath at temperature $T=D/k_B$, of a
paradigmatic model of long-range interactions, the Hamiltonian
mean-field model \cite{Ruffo:1995}. Here, $k_B$ is the Boltzmann
constant. Note that here we have set the damping
coefficient to unity.

In terms of the so-called complex order parameter
$r\exp(\mathrm{i}\psi)$ (with real $r$ and $\psi$ satisfying $0 \le r \le 1$ and
$\psi \in [0,2\pi)$), defined as 
\begin{equation}
r(t)\exp(\mathrm{i}\psi(t))\equiv
\frac{1}{N}\sum_{j=1}^N \exp(\mathrm{i}\theta_j(t)),
\end{equation}
the equations of motion (\ref{eq:Kuramoto-eom}) read 
\begin{equation}
\frac{{\rm d}\theta_i}{{\rm d}t}=\omega_i + K r(t)
\sin(\psi(t)-\theta_i)+\eta_i(t),
\end{equation}
which makes it evident the mean-field nature of the dynamics: every
oscillator evolves in a mean field of magnitude $r$ produced by all the
oscillators. The quantities $r(t)$ and $\psi(t)$ are obtained as
\begin{equation}
r(t)=\sqrt{r_x^2(t)+r_y^2(t)},~\psi(t)=\tan^{-1}(r_y(t)/r_x(t)),
\end{equation}
where
 we have 
 \begin{equation}
 (r_x(t),r_y(t))\equiv \frac{1}{N}\sum_{j=1}^N(\cos \theta_j(t),\sin
\theta_j(t)).
\end{equation}

In the thermodynamic limit $N \to \infty$, one may characterize the
dynamics (\ref{eq:Kuramoto-eom}) in terms of a single-oscillator
probability density $f(\theta,\omega,t)$ defined such that
$f(\theta,\omega,t){\rm d}\theta$ gives out of oscillators that have
natural frequency equal to $\omega$ the fraction that have their phase
values in
$[\theta,\theta+{\rm d}\theta]$ at time $t$. The function
$f(\theta,\omega,t)$ satisfies
$f(\theta+2\pi,\omega,t)=f(\theta,\omega,t)~\forall~\omega,t$ and the
normalization 
\begin{equation}
\int_0^{2\pi}{\rm
d}\theta~f(\theta,\omega,t)=1~\forall~\omega,t.
\end{equation}
The time evolution of
$f$ obeys the Fokker-Planck equation \cite{Sakaguchi:1988} 
\begin{equation}
\frac{\partial f}{\partial t}=-\frac{\partial}{\partial \theta}\left[\left(\omega+Kr
\sin(\psi-\theta)\right)f\right]+D\frac{\partial^2 f}{\partial \theta^2},
\end{equation}
with 
\begin{equation}
r(t)\exp(\mathrm{i}\psi(t))=\int_{-\infty}^\infty {\rm
d}\omega~g(\omega)\int_0^{2\pi}{\rm
d}\theta~\exp(\mathrm{i}\theta)f(\theta,\omega,t).
\end{equation}

In the limit $t\to \infty$, the dynamics (\ref{eq:Kuramoto-eom}) settles into a stationary
state. Correspondingly, the single-oscillator density
$f(\theta,\omega,t)$ assumes a time-independent form $f_{\rm
st}(\theta,\omega)$. Concomitantly, the quantities $r$ and $\psi$
assume the time-independent values $r_{\rm st}$ and $\psi_{\rm st}$,
respectively. It may be shown that only under conditions mentioned above that
reduce the set of equations (\ref{eq:Kuramoto-eom}) to (\ref{eq:BMF-eom}) does the dynamics satisfy
detailed balance. In this case, the dynamics settles
into a stationary state that is in equilibrium \cite{Gupta:2017}. Otherwise, the dynamics
(\ref{eq:Kuramoto-eom}) relaxes into a generic nonequilibrium stationary
state (NESS) \cite{Gupta:2017,Livi:2017}. In the stationary state, measuring $\theta_i$'s with respect to
$\psi_{\rm st}$ (thus setting $r_{{\rm st},y}=0, r_{\rm st}=r_{{\rm
st},x}$), we obtain the stationary-state dynamics as
\begin{equation}
\frac{{\rm d}\theta_i}{{\rm d}t}=\omega_i - K r_{\rm st} \sin
\theta_i+\eta_i(t).
\label{eq:Kuramoto-eom-stationary}
\end{equation}
In the stationary state, it is known that for given $g(\omega)$ and $K$ and on tuning the noise
strength $D$, one observes a continuous phase transition from a low-$D$
synchronized ($r_{\rm st}\ne0$) to a high-$D$ incoherent ($r_{\rm st}=0$) phase at the critical value $D_c$ that solves the equation
\cite{Sakaguchi:1988}
\begin{equation}
K=2\left[\int_{-\infty}^\infty {\rm
d}\omega~\frac{g(\omega)D_c}{D_c^2+\omega^2}\right]^{-1}.
\label{eq:Dc}
\end{equation}
In particular, for the BMF model, one obtains by substituting
$g(\omega)=\delta(\omega)$ in the above equation the critical noise
strength as $D_c=K/2$ \cite{Campa:2014}. In both the noisy Kuramoto and
the BMF model, the quantity $r_{\rm st}$
decreases continuously from the value of unity at $D=0$ to zero at
$D=D_c$, and remains zero at higher values of $D$.

It is our aim in this paper to characterize in the
thermodynamic limit the stationary-state dynamics
(\ref{eq:Kuramoto-eom-stationary}) in
terms of autocorrelations of the dynamical variable $\cos \theta$. The
reason behind choosing such a dynamical variable is that the stationary
order parameter is indeed $r_{\rm st}=\langle \cos \theta \rangle$, and
so through such a choice, we will be monitoring the correlation between
the phase coherence at two times in the stationary state. To
this end, we define the autocorrelation function as 
\begin{eqnarray}
C(\tau)&\equiv& \lim_{t\to \infty, \tau={\rm
finite}}\langle \cos \theta(t)\cos \theta(t+\tau)\rangle,
\label{eq:autocorrelation-definition}
\end{eqnarray}
which is a measure of similarity of $\cos \theta$-values
at two different times. Namely, $C(\tau)$ is a measure of the possibility of
observing a given $\cos \theta$-value (which could be contributed by any
of the $N$ oscillators) at one time and another given value at another
time. Consequently, $C(\tau)$ will be given by an appropriate joint probability, which
according to standard notions of probability theory~\cite{Papoulis:1965} may be expressed in terms of a
conditional probability.

In order to proceed, we show the similarity of the stationary dynamics
(\ref{eq:Kuramoto-eom-stationary}) with that of a single
non-uniform oscillator of frequency $\omega$~\cite{Strogatz:2014}. Contrary to an uniform oscillator for which the phase changes uniformly
in time, a non-uniform oscillator is one in which the phase has a
non-uniform variation in time: sometimes it speeds up and sometimes it
slows down. Let us then first describe
a single non-uniform oscillator characterized by its phase
$\theta \in [0,2\pi)$, whose time evolution in presence of a Gaussian,
white noise $\eta(t)$ is given by the following Langevin dynamics:
\begin{equation}
\frac{{\rm d}\theta}{{\rm d}t}=\omega-{\cal K}\sin \theta+\eta(t).
\label{eq:eom}
\end{equation}
Here, $\omega$ and ${\cal K} \ge 0$ are real constants, while $\eta(t)$
satisfies 
\begin{equation}
\langle \eta(t)\rangle=0,~\langle
\eta(t)\eta(t')\rangle=2D\delta(t-t').
\end{equation}
For $\omega=0$, the equation of
motion (\ref{eq:eom}) corresponds to overdamped dynamics of $\theta$ in 
a potential $V(\theta)\equiv -{\cal K}\cos \theta$ (with the damping
coefficient set to unity) and in contact with a
heat bath at temperature $T=D/k_B$. 

In the absence of noise (i.e., with $D=0$), and provided we have ${\cal K} > \omega$, the dynamics
(\ref{eq:eom}) has two fixed points given by
\cite{Strogatz:2014}
\begin{equation}
\overline{\theta}=\sin^{-1}(\omega/{\cal K});~
\cos \overline{\theta}=\pm \sqrt{1-\omega^2/{\cal K}^2}.
\end{equation}
In
order to determine which of the two fixed points is linearly stable, we may
linearize Eq. (\ref{eq:eom}) about $\overline{\theta}$, by writing $\theta$
as $\theta=\overline{\theta}+\delta \theta$, with $|\delta \theta|\ll 1$. The
linearized equation reads ${\rm d}\delta \theta/{\rm d}t=-{\cal K}\delta \theta
\cos \overline{\theta}$, from which it is evident that the fixed point that satisfies $\cos
\overline{\theta} >0$ is linearly stable, while the other one is
linearly unstable. Denoting the stable fixed point by $\theta_{\rm
stable}$,
we have $\theta_{\rm stable}=\sin^{-1}(\omega/{\cal K});~\cos \theta_{\rm
stable}=\sqrt{1-\omega^2/{\cal K}^2}$. In the long-time limit, the dynamics
(\ref{eq:eom}) in absence of noise results in $\theta$ settling into
the fixed-point value $\theta_{\rm stable}$. In presence of weak noise ($D
\to 0$), we expect
$\theta$ in the long-time limit to have a (narrow) distribution of values around
$\theta_{\rm stable}$.

To characterize the behavior of the dynamics (\ref{eq:eom}) in the
stationary state, attained as $t \to \infty$, let us introduce the
quantity ${\cal P}(\theta,t)$ as a one-time probability density defined such that
${\cal P}(\theta,t){\rm d}\theta$ gives the probability to observe a value of
the phase in the interval $[\theta,\theta+{\rm d}\theta]$ at time $t$.
One has the normalization $\int_0^{2\pi}{\rm
d}\theta~{\cal P}(\theta,t)=1~\forall~t$; moreover, ${\cal P}(\theta,t)$
is $2\pi$-periodic in $\theta$: ${\cal
P}(\theta+2\pi,t)={\cal P}(\theta,t)$. For a given initial condition ${\cal
P}(\theta,t=0)$, the quantity ${\cal P}(\theta,t)$ evolves in time
according to a Fokker-Planck equation
that one may derive from Eq. (\ref{eq:eom}) using standard procedure
\cite{Risken:1996}.
The equation reads
\begin{equation}
\frac{\partial {\cal P}(\theta,t)}{\partial t}=-\frac{\partial }{\partial
\theta}\left[(\omega - {\cal K} \sin \theta){\cal P}(\theta,t)\right]+D\frac{\partial^2
{\cal P}(\theta,t)}{\partial \theta^2}.
\label{eq:Fokker-Planck-equation}
\end{equation}

Comparing the dynamics (\ref{eq:Kuramoto-eom-stationary}) and
(\ref{eq:eom}), we arrive at the following useful analogy between the
system of Kuramoto oscillators and a single non-uniform oscillator.
First, let us club the Kuramoto oscillators into groups that have the same natural
frequency $\omega$. Then, the stationary-state dynamics of oscillators within each
group is that of a non-uniform oscillator, Eq. (\ref{eq:eom}), with
the constant ${\cal K}$ equal to $Kr_{\rm st}$, where $r_{\rm
st}$ is the global order parameter obtained from the stationary-state
dynamics of oscillators across all groups. This analogy will be used in
this paper to obtain results for the
noisy Kuramoto model based on those for the single oscillator. The various steps of analysis would be (i) derive
the conditional probability in the stationary state of the single
non-uniform oscillator to observe given values of the phase at two
different times, and (ii) use the derived results and the mapping
between the single non-uniform oscillator and the Kuramoto oscillators
mentioned above to obtain the stationary state conditional probability
for the latter that will be required to evaluate
(\ref{eq:autocorrelation-definition}). The method we employ to
derive our results for the single non-uniform oscillator is based on a
study of a Fokker-Planck equation of the form of Eq.
(\ref{eq:Fokker-Planck-equation}) satisfied by the conditional
probability. A general reference that summarizes techniques required to
study such equations with the help of a Fourier expansion is the book by Risken~\cite{Risken:1996}.

The paper is structured as follows. 
In Section \ref{sec:model-with-noise}, we obtain for the single
non-uniform oscillator exact analytical results for
the autocorrelation. We then use these results in
Section \ref{sec:noisy-Kuramoto-model} to derive the core results of the paper, namely, the
autocorrelation $C(\tau)$ in the stationary state of the noisy
Kuramoto model and the BMF model. We also compare our analytical results with
those obtained from direct numerical integration of the dynamical
equations of motion, demonstrating a very good agreement. The paper ends
with conclusions.

\section{The nonequilibrium stationary state of the single
non-uniform oscillator}
\label{sec:model-with-noise}

In this section, we study in detail the stationary state of the single
non-uniform oscillator. We start with obtaining the form of the probability density ${\cal P}(\theta,t)$ in the stationary state. As $t\to \infty$, one expects ${\cal P}(\theta,t)$ to relax to
a stationary distribution ${\cal P}_{\rm st}(\theta)$ that from Eq.
(\ref{eq:Fokker-Planck-equation}) is seen to satisfy 
\begin{equation}
0=-\frac{\partial}{\partial \theta}\left[(\omega - {\cal K} \sin \theta){\cal
P}_{\rm st}(\theta)\right]+D\frac{\partial^2{\cal P}_{\rm st}(\theta)}{\partial \theta^2},
\end{equation}
thereby implying that  
\begin{equation}
-D\frac{\partial {\cal P}_{\rm st}(\theta)}{\partial \theta}+(\omega -
{\cal K}
\sin \theta){\cal P}_{\rm st}(\theta)=J_{\rm st}.
\end{equation}
Here, $J_{\rm st}$, a constant independent of $\theta$, is the current in the stationary state, with
the first and the second term on the left hand side accounting for the
contribution due to diffusion and drift, respectively. This last
equation is solved easily, with the value of $J_{\rm st}$ fixed by
accounting for the $2\pi$-periodicity of ${\cal P}_{\rm st}(\theta)$. One gets
\begin{equation}
{\cal P}_{\rm st}(\theta)={\cal C}(\omega)e^{\left(-{\cal K}+{\cal K}\cos \theta+\omega
\theta\right)/D}\left[1+\frac{(e^{-2\pi \omega/D}-1)\int_0^\theta
{\rm d}\theta'~ e^{(-\omega \theta'-{\cal K} \cos \theta')/D}}{\int_0^{2\pi}{\rm
d}\theta'~e^{(-\omega \theta'-{\cal K} \cos \theta')/D}}\right],
\label{eq:stationary-distribution}
\end{equation}
where ${\cal C}(\omega)$ is a constant whose value may be fixed by employing the
normalization condition: $\int_0^{2\pi}{\rm d}\theta~{\cal P}_{\rm
st}(\theta)=1$. From Eq. (\ref{eq:stationary-distribution}), it may
be checked that for $\omega=0$, one has an equilibrium stationary state: 
\begin{equation}
{\cal P}_{\rm st}(\theta)\propto \exp\left[(-{\cal K}+{\cal K} \cos
\theta)/D\right]\sim \exp[-V(\theta)/D];
\end{equation}
substituting in the equation for $J_{\rm st}(\theta)$ given above, one
finds that the current is identically zero for all $\theta$, as it should be in equilibrium.

\subsection{Stationary correlations}
\label{sec:quantities-of-interest}

For the dynamics (\ref{eq:eom}), let $P(\theta,t|\theta',t')$ denote a
conditional probability density. Namely,
the quantity $P(\theta,t|\theta',t'){\rm d}\theta$ gives the
probability that the phase has a value in the interval
$[\theta,\theta+{\rm d}\theta]$ at time $t$, given
that it had the value $\theta'$ at an earlier time $t'\le t$. The function
$P(\theta,t|\theta',t')$ is $2\pi$-periodic in both $\theta$ and
$\theta'$:
\begin{equation}
P(\theta+2\pi,t|\theta'+2\pi,t')=P(\theta,t|\theta',t'),
\end{equation}
and satisfies the
normalization condition 
\begin{equation}
\int_0^{2\pi} {\rm
d}\theta~P(\theta,t|\theta',t')=1~\forall~\theta',t' {\rm~and~} \forall~t \ge t'.
\end{equation}
The time evolution of $P(\theta,t|\theta',t')$ follows a Fokker-Planck
equation that may be written down by using the Langevin
equation (\ref{eq:eom}). The equation is given by
\begin{equation}
\frac{\partial P(\theta,t|\theta',t')}{\partial
t}=-\frac{\partial }{\partial \theta}\left[(\omega-{\cal K}\sin
\theta)P(\theta,t|\theta',t')\right]+D\frac{\partial^2
P(\theta,t|\theta',t')}{\partial \theta^2}.
\label{eq:Fokker-Planck}
\end{equation}
Since $P(\theta,t|\theta',t')$ is $2\pi$-periodic in $\theta$ and
$\theta'$, we may
expand $P$ in a Fourier series, as 
\begin{equation}
P(\theta,t|\theta',t')=\sum_{n,m=-\infty}^\infty
\widetilde{P}_{n,m}(t|t')e^{i (n\theta+m\theta')},
\end{equation}
where the Fourier coefficients
are given by
\begin{equation}
\widetilde{P}_{n,m}(t|t')=\frac{1}{(4\pi)^2}\int_0^{2\pi} {\rm d}\theta
\int_0^{2\pi} {\rm
d}\theta'~P(\theta,t|\theta',t')e^{-i(n\theta+m\theta')}.
\end{equation}
Since $P(\theta,t|\theta',t)$ is real, we have
$\widetilde{P}_{-n,-m}(t|t')=\widetilde{P}^\star_{n,m}(t|t')$, where $\star$
denotes complex conjugation; also,
$P(\theta,t'|\theta',t')=\delta(\theta-\theta')$ implies that 
\begin{equation}
\widetilde{P}_{n,m}(t'|t')=\frac{\delta_{n,-m}}{2\pi}.
\end{equation}

Using Eq. (\ref{eq:stationary-distribution}) and the Fourier
expansion for $P(\theta,t|\theta',t')$, we
obtain the stationary correlation $C(\tau,\omega)\equiv\lim_{t'\to \infty,\tau={\rm
finite}}\langle \cos \theta(t')\cos \theta(t'+\tau)\rangle$ as
\begin{eqnarray}
\fl C(\tau,\omega)=\lim_{t'\to \infty,\tau={\rm
finite}}\int_0^{2\pi}{\rm d}\theta \int_0^{2\pi}{\rm d}\theta'~\cos \theta
\cos \theta' P(\theta,t\equiv t'+\tau|\theta',t') {\cal P}_{\rm st}(\theta') \nonumber
\\
\fl =\lim_{t'\to \infty,\tau={\rm
finite}}\pi \sum_{m=-\infty}^\infty\Big[
\left(\widetilde{P}_{1,m}(t|t')+\widetilde{P}_{-1,m}(t|t')\right)\int_0^{2\pi}{\rm
d}\theta~e^{im\theta}\cos
\theta~{\cal P}_{\rm st}(\theta)\Big]. 
\label{eq:Ctau-general-expression}
\end{eqnarray}

Now, Eq. (\ref{eq:Fokker-Planck}) gives the time
evolution of $\widetilde{P}_{n,m}(t|t')$ as
\begin{equation}
\frac{\partial \widetilde{P}_{n,m}(t|t')}{\partial
t}=-(in\omega+Dn^2)\widetilde{P}_{n,m}(t|t')+\frac{n{\cal K}}{2}\left(\widetilde{P}_{n-1,m}(t|t')-\widetilde{P}_{n+1,m}(t|t')\right).
\label{eq:eom-Fourier-coefficients}
\end{equation}
It then follows that $\widetilde{P}_{0,m}(t|t')$ is independent of $t$,
and thus, we have
\begin{equation}
\widetilde{P}_{0,m}(t|t')=\widetilde{P}_{0,m}(t'|t')=\frac{\delta_{0,-m}}{2\pi}.
\end{equation}
Let us consider separately the cases ${\cal K}\ne 0$ and ${\cal K}=0$.
\subsubsection{The case ${\cal K} \ne 0$}
\label{sec:Kne0}

For any pair of values $(n,m)$, the system of equations (\ref{eq:eom-Fourier-coefficients}) is not
closed and in fact involves an infinite hierarchy: to solve for
$\widetilde{P}_{n,m}$ requires knowing $\widetilde{P}_{n+1,m}$ whose
solution in turn requires knowing $\widetilde{P}_{n+2,m}$, and so on. 
Nevertheless, noting that for a given $m$, only one of the
$\widetilde{P}_{n,m}$'s is non-zero at the initial time, that is, $\widetilde{P}_{n,m}(t'|t')=\delta_{n,-m}/(2\pi)$, the system of
equations is solved quite easily by truncating it at a given value $n=n_{\rm
max}$, that is, by stipulating that $\widetilde{P}_{n,m}(t|t')=0$ for $n > n_{\rm max}$, 
where $n_{\rm max}$ may be chosen to be as large as possible. In
practice, in evaluating the
correlation~(\ref{eq:Ctau-general-expression}), we choose the same
$n_{\rm max}$ for different $m$'s and restrict the values of $m$ to the
range $[-n_{\rm max}:n_{\rm max}]$, checking that a larger value of
$n_{\rm max}$ does not lead to any significant change in the results obtained. 

From the system of equations
(\ref{eq:eom-Fourier-coefficients}), one may obtain closed form
expressions for $\widetilde{P}_{n,m}$ for the particular case of small
${\cal K}$, when the equations can be solved perturbatively. To this end, we expand
$\widetilde{P}_{n,m}(t|t')$ as a power series in ${\cal K}$, as
\begin{equation}
\widetilde{P}_{n,m}(t|t')=\widetilde{P}^{(0)}_{n,m}(t|t')+{\cal
K}\widetilde{P}^{(1)}_{n,m}(t|t')+{\cal K}^2\widetilde{P}^{(2)}_{n,m}(t|t')+\ldots,
\end{equation}
where we have $\widetilde{P}^{(\alpha)}_{n,m}(t|t')=O({\cal K}^0)$ for $\alpha \ge 1$.
Substituting in Eq. (\ref{eq:eom-Fourier-coefficients}), and
comparing terms of the same order in ${\cal K}$ from both sides, we get
\begin{eqnarray}
\fl {\cal K}^0: \frac{\partial \widetilde{P}^{(0)}_{n,m}(t|t')}{\partial
t}=-(in\omega+Dn^2)\widetilde{P}^{(0)}_{n,m}(t|t'),
\label{eq:hierarchy-1} \\
\fl {\cal K}^1: \frac{\partial \widetilde{P}^{(1)}_{n,m}(t|t')}{\partial
t}=-(in\omega+Dn^2)\widetilde{P}^{(1)}_{n,m}(t|t')+\frac{n}{2}\left(\widetilde{P}^{(0)}_{n-1,m}(t|t')-\widetilde{P}^{(0)}_{n+1,m}(t|t')\right),
\label{eq:hierarchy-2}
\end{eqnarray}
and so on. Since $\widetilde{P}_{n,m}(t'|t')=\delta_{n,-m}/(2\pi)$, we may take
\begin{equation}
\widetilde{P}^{(0)}_{n,m}(t'|t')=\frac{\delta_{n,-m}}{2\pi},~\widetilde{P}^{(\alpha)}_{n,m}(t'|t')=0~\forall~\alpha
\ge 1.
\end{equation}
Solving Eq. (\ref{eq:hierarchy-1}), we get
\begin{equation}
\widetilde{P}^{(0)}_{n,m}(t|t')=\frac{\delta_{n,-m}}{(2\pi)}\exp\left[-\left(in\omega+Dn^2\right)(t-t')\right],
\end{equation}
which when used in Eq. (\ref{eq:hierarchy-2}) yields the solution
\begin{eqnarray}
&&\widetilde{P}^{(1)}_{n,m}(t|t')=\frac{ne^{-(in\omega+Dn^2)(t-t')}}{4\pi}\Big[\frac{\delta_{n-1,-m}}{2nD-D+i\omega}\left(e^{-(-2nD+D-i\omega)(t-t')}-1\right)\nonumber
\\
        &&+\frac{\delta_{n+1,-m}}{2nD+D+i\omega}\left(e^{-(2nD+D+i\omega)(t-t')}-1\right)\Big].
\end{eqnarray}
To order ${\cal K}$, substituting
\begin{equation}
\widetilde{P}_{n,m}(t|t')=\widetilde{P}^{(0)}_{n,m}(t|t')+{\cal
K}\widetilde{P}_{n,m}^{(1)}(t|t')
\end{equation}
in Eq. (\ref{eq:Ctau-general-expression})
allows to obtain $C(\tau,\omega)$ for the model (\ref{eq:eom}) in the limit
of small ${\cal K}$, as 
\begin{eqnarray}
\fl C(\tau,\omega)=\Big[\frac{1}{2}e^{-(D+i\omega)\tau}\int_0^{2\pi} {\rm
d}\theta~\cos \theta e^{-i\theta}{\cal P}_{\rm st}(\theta)+
\frac{{\cal K}}{4}\Big\{\frac{(1-e^{-(D+i\omega)\tau})}{(D+i\omega)}\int_0^{2\pi}{\rm
d}\theta~\cos \theta{\cal P}_{\rm st}(\theta) \nonumber \\ 
\fl
+\frac{\left(e^{-(4D+2i\omega)\tau}-e^{-(D+i\omega)\tau}\right)}{3D+i\omega}\int_0^{2\pi} {\rm
d}\theta~\cos \theta e^{-i2\theta}{\cal P}_{\rm st}(\theta)\Big\}\Big]+{\rm c.c.},
\label{eq:Ctau-smallK-expression}
\end{eqnarray}
where ${\rm c.c.}$ denotes complex conjugate of the bracketed term.

For $\omega=0$, when the dynamics (\ref{eq:eom}) has an equilibrium
stationary state 
\begin{equation}
{\cal P}_{\rm st}(\theta)=\exp[-V(\theta)/D]/\int_0^{2\pi}{\rm d}\theta
\exp\left[-V(\theta)/D\right],
\end{equation}thereby
implying that 
\begin{equation}
\int_0^{2\pi}{\rm d}\theta~\cos \theta \sin(m\theta){\cal
P}_{\rm st}(\theta)=0
\end{equation}
for non-zero integer $m$, we obtain from Eq.
(\ref{eq:Ctau-smallK-expression}) the equilibrium correlation for small
${\cal K}$ as
\begin{eqnarray}
\fl C(\tau,\omega)\Big|_{\omega=0}=e^{-D\tau}\int_0^{2\pi}{\rm
d}\theta~\cos^2\theta\,{\cal P}_{\rm st}(\theta)\nonumber \\
\fl +\frac{{\cal
K}}{2}\Big\{\frac{(1-e^{-D\tau})}{D}\int_0^{2\pi}{\rm d}\theta~\cos \theta
\,{\cal P}_{\rm st}(\theta)+\frac{\left(e^{-4D\tau}-e^{-D\tau}\right)}{3D}\int_0^{2\pi}{\rm
d}\theta~\cos \theta \cos (2\theta)\,{\cal P}_{\rm st}(\theta)\Big\}.
\nonumber \\
\label{eq:correlation-BMF-smallK}
\end{eqnarray}
\subsubsection{The case ${\cal K}=0$}
\label{sec:K0}

For ${\cal K}=0$, Eq. (\ref{eq:stationary-distribution}) gives
${\cal P}_{\rm st}(\theta)=1/(2\pi)$, while Eq. (\ref{eq:eom-Fourier-coefficients}) with the initial condition
$\widetilde{P}_{n,m}(t'|t')=\delta_{n,-m}/(2\pi)$ has the solution
\begin{equation}
\widetilde{P}_{n,m}(t|t')=\frac{\delta_{n,-m}}{(2\pi)}\exp(\left[-\left(in\omega+Dn^2\right)(t-t')\right]).
\end{equation}
Equation (\ref{eq:Ctau-general-expression}) then yields the exact result
\begin{equation}
C(\tau,\omega)=\frac{1}{2}\cos(\omega \tau)e^{-D\tau}.
\label{eq:correlation-K0}
\end{equation}
We thus obtain $C(0)=\langle \cos^2 \theta\rangle_{\rm st}=1/2$. Putting $\omega=0$,
Eq. (\ref{eq:correlation-K0}) gives the equilibrium correlation as
\begin{equation}
C(\tau,\omega)|_{\omega=0}=\frac{1}{2}\exp(-D\tau).
\end{equation}

\section{Results for the noisy Kuramoto model}
\label{sec:noisy-Kuramoto-model}

In order to derive our results for the stationary state of the noisy
Kuramoto model, Eq. (\ref{eq:Kuramoto-eom-stationary}), we employ
the aforementioned exact analogy that exists between it and the
dynamics (\ref{eq:eom}) of the single non-uniform oscillator on
setting the constant ${\cal K}$ in the latter to the value $Kr_{\rm
st}$, where $K$ is as usual the coupling constant of the Kuramoto model
and $r_{\rm st}$ is the stationary Kuramoto order parameter.
Consequently, we may
write down for the noisy Kuramoto model the single-oscillator probability density in
the stationary state by using Eq. (\ref{eq:stationary-distribution}) as \cite{Sakaguchi:1988}
\begin{eqnarray}
\fl f_{\rm st}(\theta,\omega)={\cal C}(\omega)e^{\left(-Kr_{\rm st}+Kr_{\rm st}\cos
\theta+\omega \theta\right)/D}\left[1+\frac{(e^{-2\pi \omega/D}-1)\int_0^\theta
{\rm d}\theta'~ e^{(-\omega \theta'-Kr_{\rm st} \cos \theta')/D}}{\int_0^{2\pi}{\rm
d}\theta'~e^{(-\omega \theta'-Kr_{\rm st} \cos \theta')/D}}\right],
\label{eq:fst}
\end{eqnarray}
with the constant ${\cal C}(\omega)$ fixed by the normalization condition
$\int_0^{2\pi}{\rm d}\theta~f_{\rm st}(\theta,\omega)=1~\forall~\omega$,
and $r_{\rm st}$ determined from the self-consistent equation
\begin{equation}
r_{\rm st}=\int_{-\infty}^\infty {\rm d}\omega~g(\omega)\int_0^{2\pi}
{\rm d}\theta~\cos \theta~f_{\rm st}(\theta,\omega).
\label{eq;rst-self-consistent}
\end{equation}
The single-oscillator $\theta$-distribution ${\cal P}_{\rm st}(\theta)$,
defined as the probability density to observe a phase value equal to
$\theta$ in the stationary state, is obtained from $f_{\rm
st}(\theta,\omega)$ as
\begin{equation}
{\cal P}_{\rm st}(\theta)=\int_{-\infty}^\infty {\rm
d}\omega~g(\omega)f_{\rm st}(\theta,\omega),
\label{eq:Pst-Kuramoto}
\end{equation}
while the stationary correlation $C(\tau)$ for the Kuramoto
oscillators is obtained as 
\begin{equation}
C(\tau)=\int_{-\infty}^\infty {\rm d}\omega~g(\omega)C(\tau,\omega).
\label{eq:Ctau}
\end{equation}
Results for the BMF model, which corresponds to the case
$g(\omega)=\delta(\omega)$, may be obtained by using Eqs. (\ref{eq:fst}), (\ref{eq;rst-self-consistent}), and
(\ref{eq:Pst-Kuramoto}), as
\begin{eqnarray}
&&{\cal P}_{\rm st}(\theta)=\frac{e^{\left(Kr_{\rm st}\cos
\theta\right)/D}}{\int_0^{2\pi} {\rm d}\theta~e^{\left(Kr_{\rm st}\cos
\theta\right)/D}};~~r_{\rm st}=\frac{I_1(r_{\rm st}/D)}{I_0(r_{\rm st}/D)},
\label{eq:Pst-rst-BMF}
\end{eqnarray}
where $I_n(x)$ is the modified Bessel function of the first kind.

For a given choice of $g(\omega)$ and a given value of $K$, the explicit steps involved in
obtaining the correlation $C(\tau)$ in the synchronized and
the incoherent phase of the Kuramoto model are as follows.
For $D \ge D_c$, when ${\cal
K}=Kr_{\rm st}=0$, Eq. (\ref{eq:correlation-K0}) gives
\begin{eqnarray}
C(\tau)=\frac{e^{-D\tau}}{2}\int_{-\infty}^\infty {\rm
d}\omega~g(\omega)\cos(\omega \tau);~~D \ge D_c.
\end{eqnarray}
For $D<D_c$, when ${\cal K}=Kr_{\rm st} \ne 0$, we first obtain the
value of $r_{\rm st}$ by solving the self-consistent equation
(\ref{eq;rst-self-consistent}). We then solve for
every value of $\omega$ in the support of $g(\omega)$ the
system of equations (\ref{eq:eom-Fourier-coefficients}) with the
substitution ${\cal K}=Kr_{\rm st}$, and use the solution $\widetilde{P}_{n,m}$ in
Eq. (\ref{eq:Ctau-general-expression}), with ${\cal P}_{\rm st}(\theta)$
given by Eq. (\ref{eq:Pst-Kuramoto}), to obtain ${\cal
C}(\tau,\omega)$. Finally, Eq. (\ref{eq:Ctau})
yields the desired correlation. For the particular case of $D \lae D_c$,
so that ${\cal K}=Kr_{\rm
st}$ is small, we may use Eq. (\ref{eq:Ctau-smallK-expression}) to get 
\begin{eqnarray}
\fl C(\tau)=\int_{-\infty}^\infty {\rm
d}\omega~g(\omega)~\Big(\Big[\frac{1}{2}e^{-(D+i\omega)\tau}\int_0^{2\pi} {\rm
d}\theta~\cos \theta~e^{-i\theta}{\cal P}_{\rm st}(\theta) \nonumber \\
\fl+ \frac{Kr_{\rm st}}{4}\Big\{\frac{(1-e^{-(D+i\omega)\tau})}{(D+i\omega)}\int_0^{2\pi}{\rm
d}\theta~\cos \theta{\cal P}_{\rm st}(\theta) \nonumber \\ 
\fl +\frac{\left(e^{-(4D+2i\omega)\tau}-e^{-(D+i\omega)\tau}\right)}{3D+i\omega}\int_0^{2\pi} {\rm
d}\theta~\cos \theta~e^{-i2\theta}{\cal P}_{\rm
st}(\theta)\Big\}\Big]+{\rm c.c.}\Big);~~D \lae D_c. 
\end{eqnarray}

The correlations in the equilibrium stationary state of the BMF model
are obtained from Eq. (\ref{eq:correlation-K0}) as 
\begin{equation}
{\cal
C}(\tau)=\exp(-D\tau)/2;~D \ge D_c,
\end{equation}
while that
for $D<D_c$ are obtained by first solving the self-consistent equation
for $r_{\rm st}$ given in Eq. (\ref{eq:Pst-rst-BMF}) and then using this value to solve the system of equations
(\ref{eq:eom-Fourier-coefficients}) with $\omega=0$, and finally using
(\ref{eq:Ctau-general-expression}) with $\omega=0$ and with ${\cal
P}_{\rm st}(\theta)$ given in Eq. (\ref{eq:Pst-rst-BMF}). In
particular, the result for $D \lae D_c$ is
obtained from Eq. (\ref{eq:correlation-BMF-smallK}) with the substitution ${\cal
K}=Kr_{\rm st}$.

\begin{figure}
\centering
\includegraphics[width=80mm]{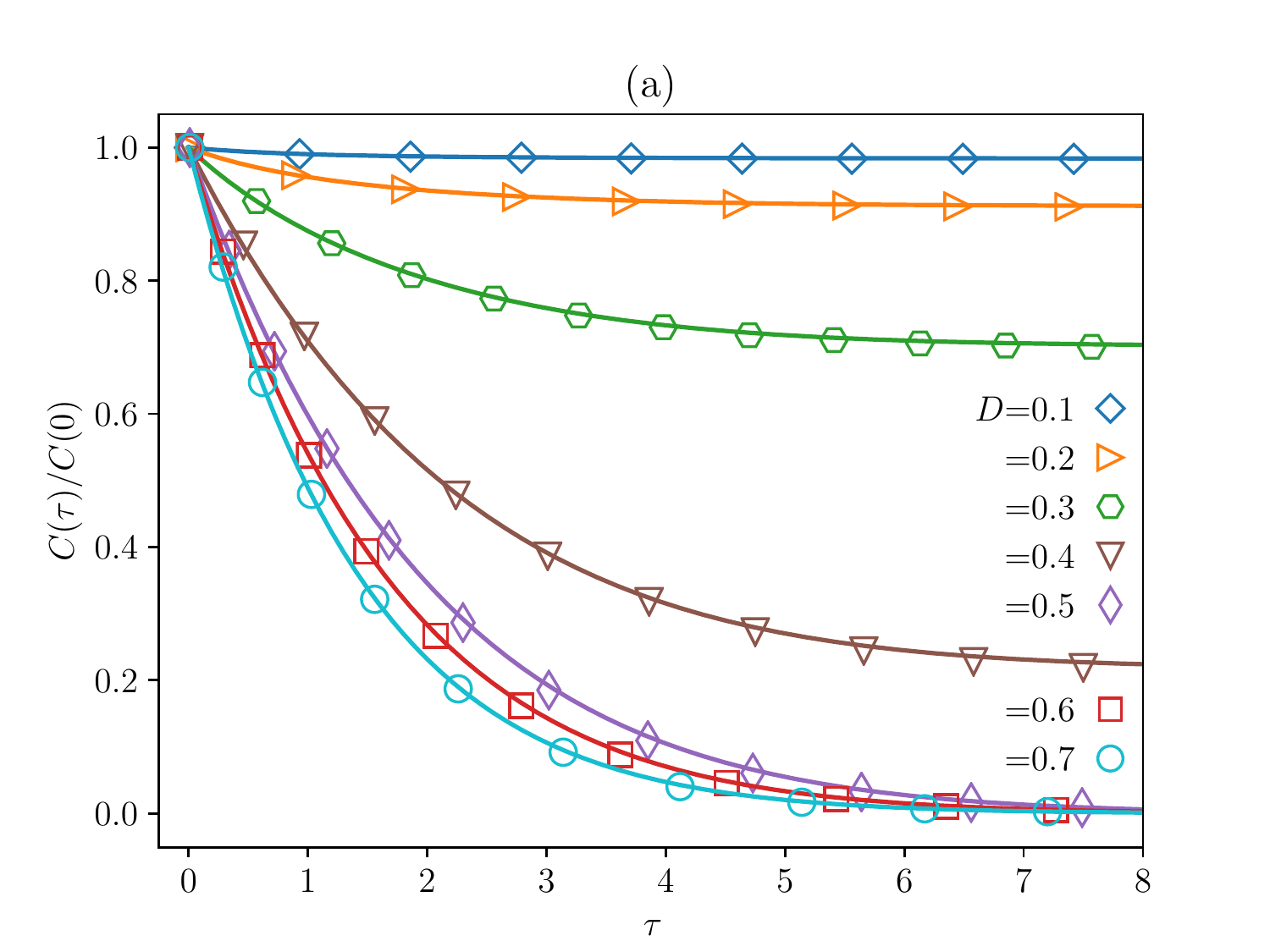}
\includegraphics[width=80mm]{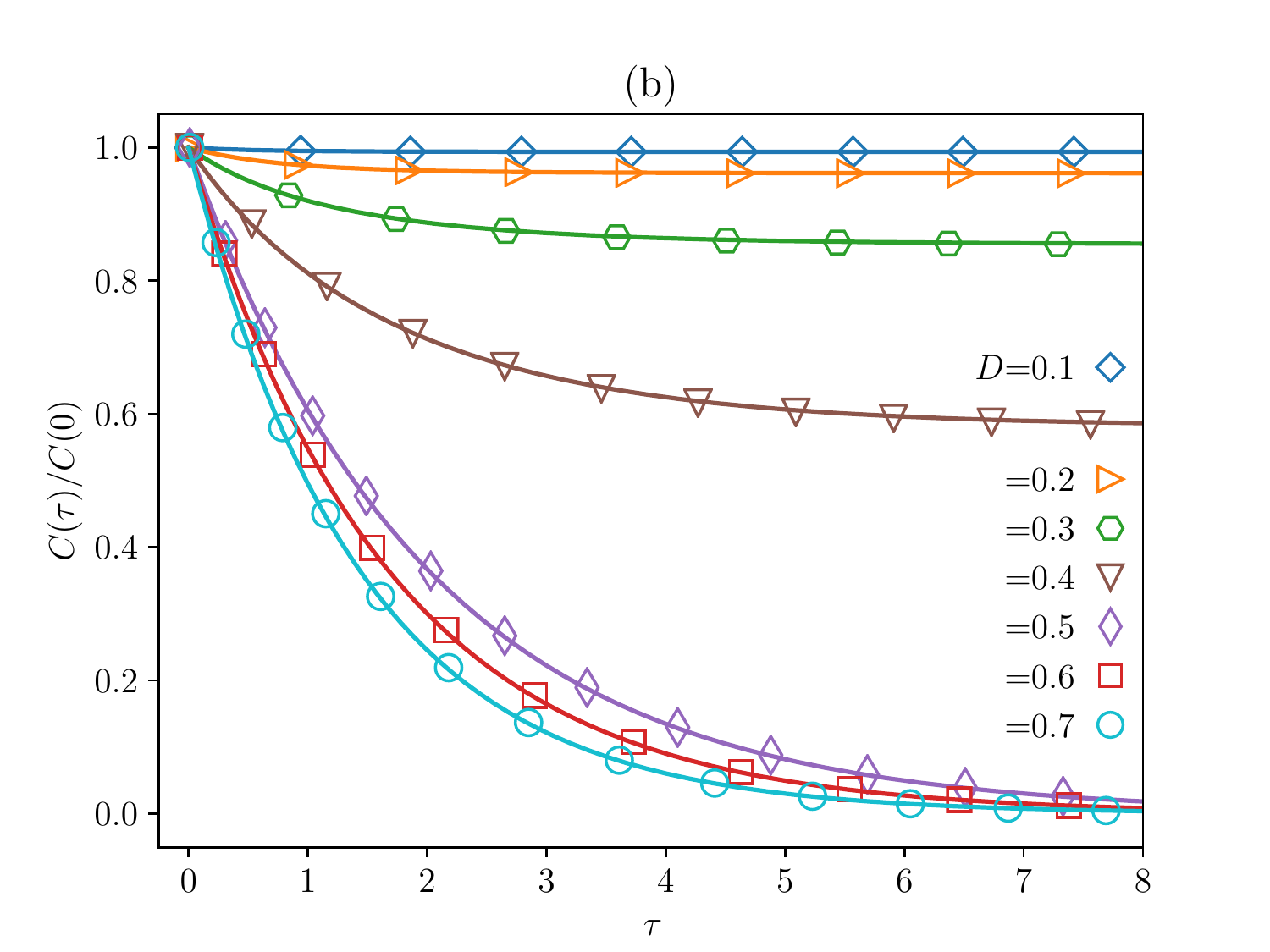}
\includegraphics[width=80mm]{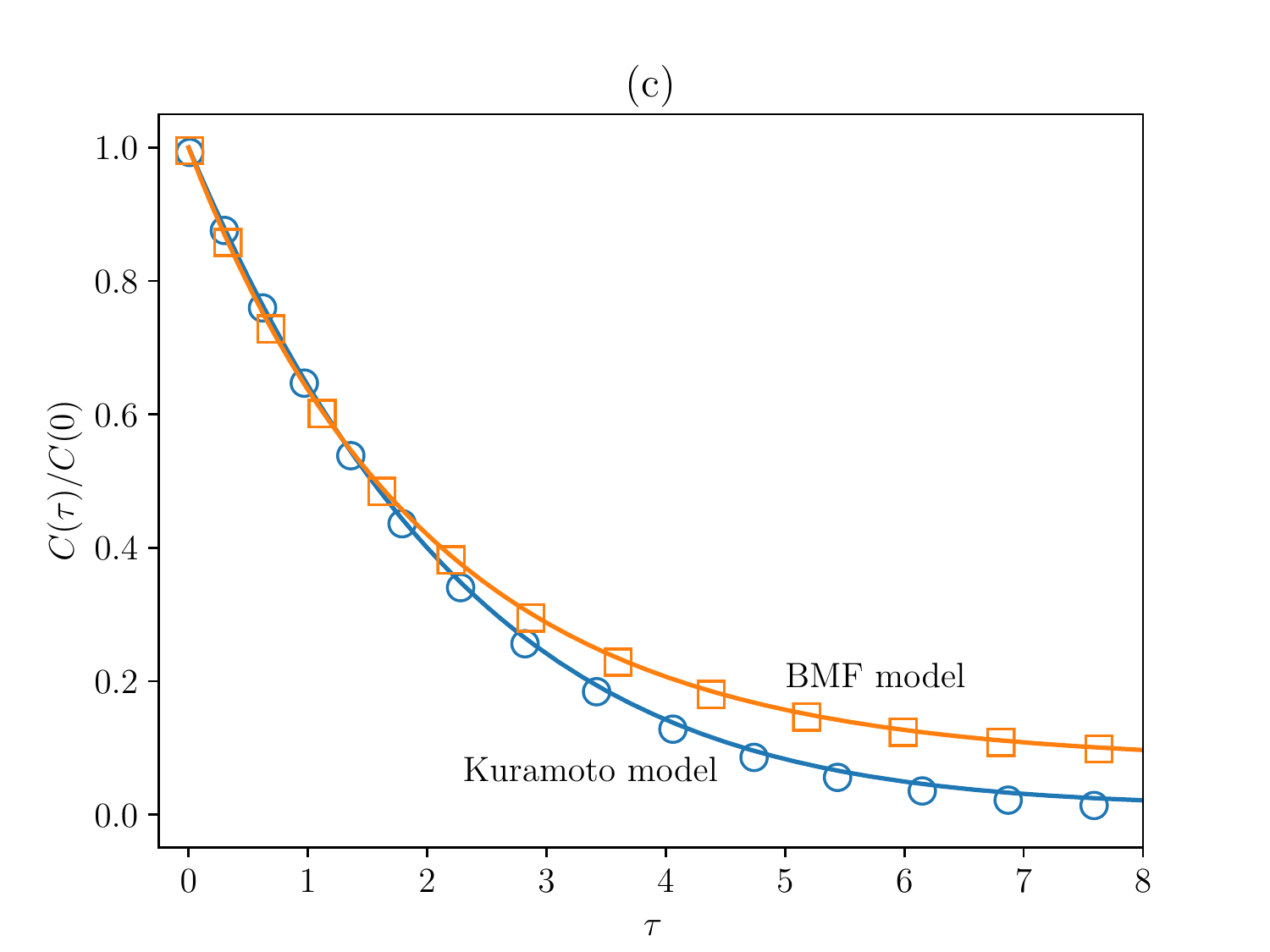}
\caption{Stationary autocorrelation $C(\tau)$, Eq. (\ref{eq:autocorrelation-definition}), normalized by its
value at $\tau=0$, in the noisy Kuramoto
(panel (a)) and the BMF model (panel (b)). For the former, the frequency
distribution is taken to be a Gaussian centered at zero and with width
given by $\sigma=0.2$, thus yielding the critical noise strength $D_c
\approx 0.43$, while that for the BMF model is given by $D_c=0.5$. In both cases, we have taken the coupling
constant to be $K=1$. Here, the points refer to results obtained from
numerical integration of the dynamics (system size: $N=10^5$ for the
noisy Kuramoto model and $N=10^6$ for the BMF
model), while continuous lines refer to
analytical results obtained in the text. In panel (c), we show for $D
\lae D_c$ and for both the models a
comparison between numerical results and theory developed in this
special case in the text; we have taken $D=0.427$ for the Kuramoto model
and $D=0.49$ for the BMF model.} 
\label{fig:Kuramoto-BMF}
\end{figure}

Following the above procedure, and considering $g(\omega)$ to be a Gaussian
distribution centered at zero and with width equal to $\sigma$, we show
in Fig. \ref{fig:Kuramoto-BMF}(a) our analytical results for the
noisy Kuramoto model compared against those obtained from direct
numerical integration of the dynamics (\ref{eq:Kuramoto-eom}) for very
large $N$ (note that all our analytical computations were done in the
limit $N \to \infty$). The
corresponding results for the BMF model are in Fig.
\ref{fig:Kuramoto-BMF}(b), while the results for $D \lae D_c$ are shown
in \ref{fig:Kuramoto-BMF}(c) for both the models. We have taken the coupling constant to be $K=1$. The function $C(\tau)$ indeed decays to the value $(r_{\rm
st})^2$ for large $\tau$, as it should. In all
cases, a very good agreement between theory and numerical
results is evident from the plots, for values of $D$ both below and above the critical value
$D_c$. The latter equals $0.5$ for the BMF model and has the value $\approx 0.43$ for the noisy Kuramoto model (obtained by numerically
solving Eq. (\ref{eq:Dc})). The autocorrelation $C(\tau)$ in all cases decays as an exponential with a rate that increases
continuously with the noise strength. An implication of this result is
that the more noisy the dynamics is, the
faster it takes to generate uncorrelated configurations.

\section{Conclusions}
\label{sec:conclusions}

In this work, we obtained exact analytical results on correlations in
the nonequilibrium stationary state of a paradigmatic many-body
interacting system showing spontaneous order, the Kuramoto model
of coupled oscillators, evolving in presence of Gaussian, white noise.
The method relies on an exact mapping of the stationary-state dynamics
of the model to the noisy dynamics of a single, non-uniform oscillator, which could
be possible due to the mean-field nature of the dynamics of the Kuramoto
model. Namely, the dynamics may be thought as that of a single
oscillator evolving in a mean-field due to its interaction with all the
other oscillators. Although we considered the Kuramoto model that
involves a sinusoidal interaction between the oscillators, the method is
designed to work for any form of interaction between the oscillators so
long as it is of the mean-field type. Investigations are underway to extract
further implications of the mapping that allow to obtain useful and physically relevant results on both static
and dynamic properties of both Kuramoto and related models. It would be interesting to obtain correlations in the stationary state of
the Kuramoto model with inertia~\cite{Gupta:2018}. As is well known, inertia significantly
changes the nature of the synchronization transition of the model, and
is thus expected to affect also the stationary-state correlation. On the
analytical side, inertia brings in significant complexity in obtaining
even the stationary state of the model~\cite{Gupta:2018}. Addressing inertial effects on autocorrelation is left for future work. 

\section{Acknowledgements}
The work of Debraj Das is supported by UGC-NET Research Fellowship Sr.
No. 2121450744, dated 29-05-2015, Ref. No. 21/12/2014(ii) EU-V.
SG acknowledges fruitful discussions with Ashik
Iqubal.

\end{document}